\documentclass[12pt]{article}
 \usepackage{graphicx}  % standard LaTeX graphics tool;
\usepackage{amsmath}   % For getting proper math eqns
\usepackage[compress]{cite}
\usepackage{amssymb}   % Used for getting various symbols eg. gtrsim, lesssim etc.
\usepackage{bm} % For bold math (esp of lower greek letters)
\usepackage{dcolumn}% Align table columns on decimal point
\usepackage{ulem}
\usepackage{color}
\usepackage{mathtools}
\usepackage{mathrsfs}
\usepackage{amsfonts}
\usepackage{varioref}
\usepackage{textcomp}
\usepackage{geometry}
\usepackage{tcolorbox}
\usepackage{empheq}
\usepackage[active]{srcltx}
\usepackage{tikz}
\usepackage{lmodern}
\usepackage{braket}
\usetikzlibrary{decorations.pathmorphing}
\tikzset{snake it/.style={decorate, decoration=snake}}
\usepackage{fancybox}
\usepackage{color}
\makeatletter
\def\mathcolor#1#{\@mathcolor{#1}}
\def\@mathcolor#1#2#3{%
	\protect\leavevmode
	\begingroup\color#1{#2}#3\endgroup
}
\RequirePackage[colorlinks,citecolor=blue,urlcolor=magenta,linkcolor=blue]{hyperref}
\allowdisplaybreaks

 \def\nn{\nonumber}
 \def\cst{\cos^2\theta}
\def\sst{\sin^2\theta}

\def\be{\begin{equation}}
\def\bea{\begin{eqnarray}}
\def\ee{\end{equation}}
\def\eea{\end{eqnarray}}

\numberwithin{equation}{section}

\newcommand{\wt}{\widetilde}

\newcommand{\diff}{\mathrm d}
\newcommand{\wf}{k_\phi}
\newcommand{\ws}{k_\psi}
\newcommand{\jr}{J_{3R}}
\newcommand{\jl}{J_{3L}}
\newcommand{\qone}{Q^{(1)}}
\newcommand{\qtwo}{Q^{(2)}}
\newcommand{\qthree}{Q^{(3)}}
\newcommand{\OO}{{\cal O}}

\def\cst{\cos^2\theta}
\def\sst{\sin^2\theta}

\def\non{\nonumber}

\newcommand{\Qi}{Q^{(I)}}
\begin{document}

\baselineskip 18pt

\thispagestyle{empty}
\begin{center}
    ~\vspace{5mm}
    
    {\Large \bf
    
Attractor saddle for 5D black hole index    }
    
    \vspace{0.4in}
    
  {
Soumya Adhikari$^1$, Pavan Dharanipragada$^2$, \\ Kaberi Goswami$^3$, Amitabh Virmani$^3$   }
\vspace{0.4in}

  {\small     $^1$School of Physics, \\ Indian Institute of Science Education and Research Thiruvananthapuram, \\ Vithura, Thiruvananthapuram, India 695551
     \vskip1ex 
    $^2$Centre for Strings, Gravitation and Cosmology, \\ Department of Physics, \\ Indian Institute of Technology Madras,  Chennai, India 600036 
     \vskip1ex
$^3$Chennai Mathematical Institute, \\ H1, SIPCOT IT Park, Siruseri, Kelambakkam, India  603103 \vspace{0.1in} }
    
    {\tt soumya12physics20@iisertvm.ac.in,
    ic40178@imail.iitm.ac.in, \\ kaberi@cmi.ac.in, avirmani@cmi.ac.in}
\end{center}

\vspace{0.4in}

 \begin{abstract}
In a recent paper, Anupam, Chowdhury, and Sen [arXiv:2308.00038] constructed the non-extremal saddle that reproduces the supersymmetric index  of the BMPV black hole with three independent charges in the classical limit. This saddle solution is a finite temperature complex solution saturating the BPS bound.  In this paper, we write this solution in a canonical form in terms of harmonic functions on three-dimensional flat base space, thereby showing that it is supersymmetric. We also show that it exhibits the new form of attraction.

\end{abstract}

\pagebreak

\baselineskip=18pt

\tableofcontents

\section{Introduction}

String theory has been very successful in  counting the microstates of a class of supersymmetric black holes \cite{Strominger:1996sh, Sen:2007qy}. In the large charge limit, the exact microscopic answer for the supersymmetric index, typically known as a Fourier coefficient of a modular form, precisely  reproduces the exponential of the Bekenstein-Hawking entropy of the corresponding supersymmetric black hole. This matching is quite remarkable, but it leaves some questions unanswered regarding the relation between the supersymmetric index and the degeneracy.

In recent years,  a method has been proposed for computing the index directly on the
gravity side\cite{2107.09062, 2011.01953}; see \cite{1810.11442} for earlier work on AdS
black holes. Using this approach,  the result of the index  computation has been shown to agree with the  Bekenstein-Hawking entropy~\cite{Boruch:2023gfn,Hegde:2023jmp,2401.13730, 2402.03297, 2306.07322,Anupam:2023yns}, including logarithmic corrections
\cite{2306.07322,Anupam:2023yns},  and very recently including higher derivative corrections \cite{2402.03297, Hegde:2024bmb}. Much of these developments have focused on four-dimensional black holes.

The aim of this paper is to further explore these ideas in five dimensions. There are 
already some key results in the literature in 5D: (i) Anupam, Chowdhury, and Sen (ACS) in ref.~\cite{Anupam:2023yns} constructed the saddle solution for the gravitational path integral that computes the index for the five-dimensional BMPV black hole with three independent charges, (ii)  ref.~\cite{Hegde:2023jmp} showed that the ACS saddle solution with equal charges is indeed supersymmetric in five-dimensional minimal supergravity, and very recently  (iii)  in ref.~\cite{Cassani:2024kjn} the authors have computed the on-shell supergravity action for  the five-dimensional saddle solution using equivariant localisation. In this paper, we focus on the five-dimensional ACS saddle solution in U(1)$^3$ supergravity with three-independent charges as constructed in~\cite{Anupam:2023yns}.

The rest of the paper is organised as follows. The ACS construction does not manifest the fact that the saddle solution is supersymmetric. In section \ref{sec:ACS_BW}, 
we write the solution in a canonical form in terms of harmonic functions on three-dimensional flat base space, thereby showing that it is supersymmetric.  In section \ref{sec:attractor}, via the 4D-5D connection, we show that the ACS saddle solution exhibits the new form of attraction \cite{Boruch:2023gfn}, i.e., the coefficients at the poles of the harmonic functions are temperature and moduli independent and are determined by the solutions to the attractor equations for the BMPV black hole charges.  We close with a brief discussion in section \ref{sec:conclusions}.

\section{The ACS saddle solution in the Bena-Warner form}
\label{sec:ACS_BW}

\label{sec:ACS_saddles}

We start with eleven-dimensional supergravity. The bosonic fields are
the metric and a 3-form potential ${\cal A}$. The bosonic Lagrangian is 
\be
{\cal L}_{11} = 
R_{11} \star_{11} 1 -
\frac{1}{2} {\cal F} \wedge \star_{11}{\cal F} - \frac{1}{6} {\cal F}
\wedge {\cal F} \wedge {\cal A},
\label{lag_11}
\ee
where $R_{11}$  denotes the eleven-dimensional Ricci
scalar, ${\cal F} = d {\cal A}$, and $\star_{11}$ is the eleven-dimensional Hodge star. We are interested in
five-dimensional theory obtained by dimensional reduction
on $T^6$ for the following form of the bosonic fields, 
\bea
ds_{11}^2 &=&  ds^2_{5} + h^1 \left(dz_1^2 +
dz_2^2 \right) + h^2 \left( dz_3^2 + dz_4^2 \right) +
h^3 \left( dz_5^2 + dz_6^2 \right), \\
{\cal A} &=& A^1 \wedge dz_1 \wedge dz_2 + A^2 \wedge dz_3 \wedge dz_4 +
A^3 \wedge dz_5 \wedge dz_6.
\label{11sol}
\eea
In writing this form, we have assumed that nothing depends on the $T^6$ coordinates. As a result, we can regard $ds^2_{5}$ as a five-dimensional metric and $h^I$ and
$A^I$ as five-dimensional scalars and vectors, respectively. 
To ensure that the five-dimensional metric is the Einstein-frame metric we also assume that the scalars $h^I$ obey the constraint
\be
\label{eqn:constr}
 h^1 h^2 h^3 = 1.
\ee
With these assumptions, the eleven-dimensional supergravity reduces to  five-dimensional U(1)$^3$ supergravity, whose bosonic Lagrangian is:
\be
 {\cal L}_5 =  
 R \star 1 - G_{IJ} dh^I \wedge
\star dh^J - G_{IJ} F^I \wedge \star F^J - \frac{1}{6} C_{IJK} F^I
\wedge F^J \wedge A^K, 
\label{lag_5}
\ee
where 
\be
G_{IJ} = \frac{1}{2}(h^I)^{-2} \delta_{IJ},
\ee
and $C_{IJK}=1$ if $(IJK)$ is a permutation of $(123)$ and $C_{IJK}=0$
otherwise.  The Maxwell field strengths are $F^I = dA^I$.

The ACS saddle solution is conjectured to be a supersymmetric solution to this five-dimensional theory. 
Building upon the seminal work of~\cite{Gauntlett:2002nw}, supersymmetric solutions to this theory have been studied in great detail~\cite{Bena:2004de, Gauntlett:2004qy, Bena:2005va}. It is well known that  a large class of supersymmetric solutions of this theory can be written in a canonical  form in terms of 8 harmonic functions on three-dimensional flat base space; see \cite{Bena:2007kg} for a review and further references. We call this canonical form the Bena-Warner form. 
A key aim of this paper is to write the ACS solution in the Bena-Warner form.

We can think of the ACS solution as the Cvetic-Youm solution \cite{Cvetic:1996xz} with certain constraints on the parameters. The Cvetic-Youm solution is labelled by six parameters: mass parameter $m$, rotation parameters $a, b$, and charge parameters $\delta_I, I = 1,2,3$. In \cite{Wu:2011gq}, Wu observed that the  Cvetic-Youm metric can be written in the following simple form,
\bea
\diff s^2 &=& ({\cal H}_1 {\cal H}_2 {\cal H}_3)^{1/3}\bigg[ -\diff t^2
+\frac{\Sigma r^2}{\bar{\Delta}_r}\, \diff r^2 +\Sigma\, \diff \theta^2
+(r^2+a^2)\sin^2\theta\, \diff \phi^2 \nonumber \\
&& +(r^2+b^2)\cos^2\theta\, \diff \psi^2
+\frac{2ms_1^2}{ {\cal H}_1 \Sigma (s_1^2-s_2^2)(s_1^2-s_3^2)}k_1^2 \nonumber \\
&& +\frac{2ms_2^2}{ {\cal H}_2 \Sigma (s_2^2-s_1^2)(s_2^2-s_3^2)}k_2^2
+\frac{2ms_3^2}{ {\cal H}_3 \Sigma (s_3^2-s_1^2)(s_3^2-s_2^2)}k_3^2 \bigg] \, , \label{CY3c}\eea
where
\bea
k_I &=&  \frac{s_Ic_1c_2c_3}{c_I}\Big(\frac{c_I^2}{c_1c_2c_3}\diff t -a\sin^2\theta\, \diff \phi
-b\cos^2\theta\, \diff \psi\Big) \nonumber \\
&& +\frac{c_Is_1s_2s_3}{s_I}\Big(b\sin^2\theta\, \diff \phi +a\cos^2\theta\, \diff \psi\Big) \, ,
\label{CY3ck}
\eea
and
\bea
\bar{\Delta}_r &=& (r^2+a^2)(r^2+b^2) -2mr^2 \, , \\
{\cal H}_I &=& 1 +\frac{2ms_I^2}{\Sigma}\, , \qquad
\Sigma = r^2 + a^2 \cos^2 \theta + b^2 \sin^2 \theta \, ,
\eea
together with $c_I = \cosh\delta_I $ and $s_I = \sinh\delta_I $. The vectors and the scalars supporting the solution take the form 
\be
A^I = \frac{2m}{ {\cal H}_I \Sigma}k_I , \qquad \qquad
h^I = \frac{ ({\cal H}_1{\cal H}_2{\cal H}_3)^{1/3}}{{\cal H}_I} .
\ee

The mass $M$, three charges $Q^{(1)}$, $Q^{(2)}$, $Q^{(3)}$ and two angular momenta
$J_\phi$, $J_\psi$ are given in terms of the parameters that enter the solution as
\begin{align}
	\label{emasses}
	& Q^{(I)}=2\, m\, c_I s_I, \\
	& M = \sqrt{m^2+(Q^{(1)})^2}+ \sqrt{m^2+(Q^{(2)})^2}+ \sqrt{m^2+(Q^{(3)})^2}, \\
	& J_\phi = 2m (a c_1 c_2 c_3 - b s_1 s_2 s_3), \\ &
	J_\psi = 2m (b c_1 c_2 c_3 - a s_1 s_2 s_3),
\end{align}
where following ACS we work in units where five-dimensional Newton's constant is 
$
G_N = \pi/4.
$ It is convenient to define, 
\bea
&&	J_{3L} =  (J_\phi-J_\psi) = 2m(a-b) 
	(c_1 c_2 c_3 +s_1 s_2 s_3)\, ,
	\\
&&	J_{3R} = (J_\phi+J_\psi) = 2m(a + b) 
	(c_1 c_2 c_3 - s_1 s_2 s_3)\,.
\eea

The ACS saddle solution is obtained by taking $m \to 0$ keeping the~\textit{charges and angular momenta} fixed. This requires, 
\bea
&& \delta_I = \frac{1}{2} \sinh^{-1} \frac{Q^{(I)}}{m}, \\
&& a-b = \frac{J_{3L}}{2m(c_1 c_2 c_3 +s_1 s_2 s_3)},\\
&& a+b = \frac{J_{3R}}{2m(c_1 c_2 c_3 -s_1 s_2 s_3)},
\eea
as we take the $m\to0$ limit. We take $Q^{(I)}$'s to be positive. In the $m\to0$ limit, $a-b$ goes to zero as 
\be
a-b \simeq \frac{J_{3L} \sqrt{m}}{\sqrt{2Q^{(1)}Q^{(2)}Q^{(3)}}},
\ee
and $a+b$ diverges as 
\be
a+b \simeq \frac{\sqrt{2} J_{3R} \sqrt{Q^{(1)}Q^{(2)}Q^{(3)}}}{\sqrt{m}(Q^{(1)}Q^{(2)}+ Q^{(2)}Q^{(3)}+ Q^{(1)}Q^{(3)})}.
\ee

In the Cvetic-Youm metric, the  outer and inner horizons $r=r_{\pm}$ are at
\begin{equation}
	r_\pm = m -\frac{1}{2}a^2 -\frac12 b^2 \pm \frac12\sqrt{(a^2-b^2)^2+4m(m-a^2-b^2)}.
\end{equation}
Since in the $m\to0$ limit $a$ and $b$ diverge, so do $r_\pm$. In order to have the horizon of the final black hole at a finite location ref.~\cite{Anupam:2023yns} defines,
\begin{equation}
	\rho^2 \equiv r^2 - \frac12 (r_+^2+r_-^2),
\end{equation}
and writes the final metric in $(t, \rho, \theta, \phi, \psi)$ coordinates with $Q^{(1)}, Q^{(2)}, Q^{(3)}, J_{3L}$, and $J_{3R}$ as the five parameters. Moreover, they restrict the parameter ranges to  $Q^{(I)}> 0$ and $J_{3L}$ real such that
\be
D:= \sqrt{4 Q^{(1)} Q^{(2)} Q^{(3)} - J_{3L}^2} > 0,  \label{restriction-1}
\ee
and $J_{3R}$ lie along the positive imaginary axis,  
\be
-iJ_{3R} > 0. \label{restriction-2}
\ee 
For concreteness we further assume that $J_{3L}  > 0$ in obtaining some of the equations below; this is not really a restriction, with minor changes $J_{3L}< 0$ case can be readily incorporated in our analysis.
 The ACS metric takes the form:
\begin{align}
	\label{emetriclimit}
	\diff s^2 &= \Delta^{1/3} \Bigg[ - \frac{\left\{\cos (2 \theta ) J_{3L} J_{3R}+2 \rho^2 Q_P\right\}^2}
	{4 \, \Delta  \, Q_P^2} \diff t^2 +\frac{4Q_P^2 \rho^2}{4 \, Q_P^2\, \rho ^4 -J_{3R}^2 \left(J_{3L}^2-4 Q_T\right)} \diff \rho^2
	+\diff \theta^2 \nonumber \\
	&-\diff \phi \diff t \, \frac{\sin ^2\theta  \left\{2 \rho ^2 Q_P \left(J_{3L}+J_{3R}\right)+\cos (2 \theta ) J_{3L} J_{3R} \left(J_{3L}+J_{3R}\right)+4 Q_T J_{3R}\right\}}{2\, \Delta \, Q_P}
	\non\\
	&- \diff \psi \diff t\, \frac{\cos ^2\theta  \left\{2 \rho ^2 Q_P \left(J_{3R}-J_{3L}\right)+\cos (2 \theta ) J_{3L} J_{3R} \left(J_{3R}-J_{3L}\right)+4 Q_T  J_{3R}\right\}}{2\,  \Delta\,  Q_P}
	\non\\ & +  \diff \phi \diff \psi \, \frac{\sin ^2\theta  \cos ^2\theta  }{2\, Q_P^3 \Delta}\times\Bigg[ Q_P 
	\Bigl\{ J_{3L}^2 Q_P^2-J_{3R}^2 Q_R
	+2 J_{3R}^2 Q_T  \left(2 \rho ^2 + Q_S\right)\Bigr\} \nonumber \\ & \hskip 2in 
	+2 Q_T  \cos (2 \theta ) J_{3L} J_{3R}^3 \Bigg]
	\non\\ 
	&+
	{\sst \over \Delta} \diff \phi^2 \Bigg[
	\rho^6 + \rho^4 \Bigg\{ \frac{(4 \cos^2 \theta - 1) J_{3L} J_{3R}}{2 Q_P}+Q_S\Bigg\} \nonumber \\
	&+ \frac{\rho^2}{4 Q_P^2} \Bigg\{ 4 \cos ^2\theta  J_{3L} J_{3R} Q_P Q_S+ \left(4 \cos ^4\theta -1\right) J_{3L}^2 J_{3R}^2+4 Q_T \sin ^2\theta  J_{3R}^2+4 Q_P^3 \Bigg\}
	\nonumber\\
	&+ \frac{1}{8 Q_P^3} \Bigg\{-2 \sin ^2\theta  Q_P^3 \left(J_{3R}^2+J_{3L}^2\right)+2 J_{3R}^2 Q_P Q_S \left(\cos (2 \theta ) J_{3L}^2 + 4 Q_T \sin ^2\theta \right)
	\nonumber\\
	&\quad +4 J_{3L} J_{3R} \Big(Q_T \sin ^2\theta  \cos (2 \theta ) J_{3R}^2+Q_P^3
	\cos ^2\theta  \Big)+8 Q_T  Q_P^3 +\cos ^2(2 \theta ) J_{3L}^3 J_{3R}^3 \Bigg\} \Bigg]
	\nonumber \\ &+
	{\cst \over \Delta} \diff \psi^2 
	\Bigg[
	\rho^6 + \rho^4 \Bigg\{ \frac{(1 - 4 \sin^2\theta) J_{3L} J_{3R}}{2 Q_P}+Q_S\Bigg\} \non \\
	&+ \frac{\rho^2}{4 Q_P^2} \Bigg\{-4 \sin ^2\theta  J_{3L} J_{3R} Q_P Q_S+(4 \sin^4\theta-1 ) J_{3L}^2 J_{3R}^2+4 Q_T \cos ^2\theta  J_{3R}^2+4 Q_P^3 \Bigg\} \non\\
	&-\frac{1}{8Q_P^3}  \Bigg\{ 2 \cos ^2\theta  Q_P^3 \left(J_{3L}^2+J_{3R}^2\right)+2 J_{3R}^2 Q_P Q_S \left(\cos (2 \theta ) J_{3L}^2-4 Q_T \cos ^2\theta \right)   \non \\
	&+4 J_{3L} J_{3R} \Big(Q_P^3\sin ^2\theta  -Q_T \cos ^2\theta  \cos (2 \theta ) J_{3R}^2\Big)+J_{3L}^3 J_{3R}^3 \cos^{2}(2\theta) -8 Q_T Q_P^3\Bigg\}
	\Bigg]\,,
\end{align}
where
\bea
Q_P &\equiv& Q^{(1)} Q^{(2)} + Q^{(2)} Q^{(3)} + Q^{(3)} Q^{(1)},  \\
Q_S &\equiv& Q^{(1)} + Q^{(2)}  + Q^{(3)} ,  \\
Q_T &\equiv& Q^{(1)}Q^{(2)} Q^{(3)}, \\ 
Q_R &\equiv&  (Q^{(1)})^2 (Q^{(2)})^2 + (Q^{(2)})^2 (Q^{(3)})^2 + (Q^{(3)})^2 (Q^{(1)})^2\, ,
\eea
and $\Delta$ is  
\be
\Delta=\frac{1}{8Q_P^3}\prod\limits_{I=1}^3[\cos(2\theta)J_{3L}J_{3R}+2(\Qi+\rho^2)Q_P].
\label{DeltaExtremal}
\ee

The matter fields supporting the solution were not given in~\cite{Anupam:2023yns}. 
The vectors and the scalars take the form, 
\bea
A^I &=& \frac{1}{\Delta_I}\Big\{2\Qi Q_P\, \diff t 
 + \zeta_I \sin^2\theta\, \diff \phi 
 + \widetilde \zeta_I \cos^2\theta\, \diff \psi\Big\},
\eea
where
\be
\zeta_I =2J_{3R}\frac{Q_T}{\Qi}-(J_{3L}+J_{3R})Q_P, \quad \quad 
\widetilde \zeta_I = 2J_{3R}\frac{Q_T}{\Qi}+(J_{3L}-J_{3R})Q_P,
\ee
\be
\Delta_I = \cos(2\theta)J_{3L}J_{3R}+2(\Qi+\rho^2)Q_P, \\
\ee
and the scalars are
\be
h^I = \frac{(\Delta_1 \Delta_2 \Delta_3)^{1/3}}{\Delta_I} .
\ee

Since $J_{3R}$ appears explicitly at several places in the metric and  other fields, the solution is \textit{not a real solution} to Lorentzian supergravity.

\subsection{The four-dimensional base metric}
Metric \eqref{emetriclimit} can be written in the form, 
\begin{equation}
	\diff s^2= -f^2(\diff t +k_\phi \diff \phi +k_\psi \diff \psi)^2+f^{-1} \diff s^2_{\mathrm{4d-base}},
	\label{5dmetric}
\end{equation}
with
\bea
	f  &=& \frac{\cos (2 \theta ) J_{3L} J_{3R}+2 \rho^2 Q_P}
	{2 \, \Delta^{1/3}  \, Q_P},\\
	\wf &=& \frac{Q_P\, \sin^2 \theta\,[4 \jr Q_T+ 2(\jr+\jl) Q_P \, \rho^2 +\jl \jr \, (\jr+\jl ) \cos(2\theta)]}{(2Q_P\, \rho^2+\jl\jr\,\cos(2\theta))^2},
\nonumber \\	\label{omegaphi}\\
	\ws &=& \frac{Q_P\, \cos^2 \theta\,[4 \jr Q_T+ 2(\jr-\jl) Q_P \, \rho^2 +\jl \jr \, (\jr-\jl ) \cos(2\theta)]}{(2Q_P\, \rho^2+\jl\jr\,\cos(2\theta))^2}.
	\nonumber \\
	\label{omegapsi}
\eea
The  four-dimensional base metric $\diff s^2_{\mathrm{4d-base}}$ is given as,
\be
\diff s^2_{\mathrm{4d-base}}= g_{\rho \rho} \diff \rho^2 + g_{\theta \theta} \diff \theta^2 + g_{\phi \phi} \diff \phi^2+ g_{\psi \psi} \diff \psi^2 + 2 g_{\phi \psi} \diff \phi \diff \psi,
\ee
where
\bea
g_{\rho \rho} &=& \rho^2 \, \frac{\rho^2 +  c \, d \, \cos 2 \theta}{\rho^4 +  d^2 \, (1-c^2)}\, ,\\
g_{\theta \theta} &=&\rho^2+  c \, d \, \cos 2\theta \,  \, , \\
g_{\phi \phi} &=&(\rho^4 + 2  \, c  \, d \,  \rho^2 \cos^2 \theta + d^2  \sin^2 \theta + c^2  \, d^2  \cos 2 \theta) \frac{\sin^2 \theta}{\rho^2+  c\, d\, \cos 2\theta}\, ,\\
g_{\psi \psi} &=&  (\rho^4 - 2  \, c  \, d \,  \rho^2 \sin^2 \theta + d^2  \cos^2 \theta - c^2  \, d^2  \cos 2 \theta) \frac{\cos^2 \theta}{\rho^2+  c\, d\, \cos 2\theta}\, ,\\
g_{\phi \psi} &=&  \frac{d^2 \sin^2 \theta \cos^2 \theta}{\rho^2 + c \, d \, \cos 2 \theta}\, ,
\eea
where we have introduced two new paramaters,
\bea	
c&=& \frac{\jl}{\sqrt{4\qone\qtwo\qthree}}, \\
d&=&\frac{\jr\sqrt{\qone\qtwo\qthree}}{\qone\qtwo+\qtwo\qthree+\qthree\qone}.
\eea
We note that because of the restrictions on the parameters \eqref{restriction-1} and \eqref{restriction-2}, $(1-c^2)> 0$ and $d^2 < 0$.  We also note that the four-dimensional base metric $\diff s^2_{\mathrm{4d-base}}$  is Ricci flat.

The base metric $\diff s^2_{\mathrm{4d-base}}$ can be written in the Gibbons-Hawking form in the same way as in \cite{Hegde:2023jmp},
\begin{equation}
	\diff s^2_{\mathrm{4d-base}} = V^{-1}(\diff z+A)^2+ V\diff s^2_{\mathrm{3d-base}},
	\label{GibbHawk}
\end{equation}
where 
\be
z = \psi - \phi, \qquad \varphi = \phi + \psi,
\ee
and the three-dimensional base metric $\diff s^2_{\mathrm{3d-base}}$ with coordinates $(\rho, \theta, \varphi)$ is flat. 
The function $V$ in equation \eqref{GibbHawk} reads,
\begin{equation}
	V=\frac{4\,(\rho^2+c \, d\, \cos 2\theta)}{\rho^4+(1-c^2) \, d^2 \, \cos^2 2\theta},
	\label{harmonicV}
\end{equation}
with the 1-form $A$ satisfying 
\begin{equation}
	\star_3 \diff A=\diff V, \label{stardV}
\end{equation}
where $\star_3$ is the Hodge star in three-dimensions with the three-dimensional base metric  $\diff s^2_{\mathrm{3d-base}}$. Our conventions are $\epsilon_{\rho \theta \varphi} > 0$. The  three-dimensional base metric takes the form,
\begin{align}
	\diff s^2_{\mathrm{3d-base}}=&\frac1{4}(\rho^4+(1-c^2) \, d^2 \cos^2 2\theta)\bigg\{\frac{\rho^2}{\rho^4+(1-c^2)\, d^2}\, \diff\rho^2+\diff \theta^2\bigg\} \nonumber\\
	&+ \frac{1}{16}(\rho^4+(1-c^2)\,d^2) \sin^2 2\theta\, \diff\varphi^2.
\end{align}

To arrive at the cartesian form for the three-dimensional base, 
\be
\diff s^2_{\mathrm{3d-base}} = \diff y_1^2 + \diff y_2^2 + \diff y_3^2,
\ee
we need to do the following coordinate transformation,
\bea
	y_1&=&\frac{1}{4} \, \sqrt{\rho^4+(1-c^2)d^2} \, \sin 2\theta \, \sin \varphi,\\
	y_2&=&\frac{1}{4} \, \sqrt{\rho^4+(1-c^2)d^2} \, \sin 2\theta\, \cos \varphi,\\
	y_3&=&\frac{1}{4} \, \rho^2 \, \cos 2\theta.
\eea
\subsection{Harmonic functions}
The full solution above can now readily be written in the Bena-Warner form in terms of 8 harmonic functions $\{V, K^I, L_I, M\}$. We quickly review the Bena-Warner form of the solution and then write the 8 harmonic functions for the ACS solution. The metric takes the form,
\be
\diff s^2 = -f^2(\diff t +k)^2+f^{-1} \diff s^2_{\mathrm{4d-base}}, \label{5d-metric}
\ee
with the four-dimensional base metric given in \eqref{GibbHawk} and \eqref{stardV}, and 
where the one-form $k$ on the four-dimensional base space is, 
\be
k=\mu(\diff z +A) +\omega. \label{omega-1}
\ee
The function $\mu$ appearing in this equation is given as, 
\be
\mu= \frac{1}{6}C_{IJK} \frac{K^IK^JK^K}{V^2}+\frac{1}{2V}K^IL_I +M,
\ee
and the three-dimensional one-form $\omega$ is given as 
\be
\star_3\diff\omega=V\diff M-M\diff V+\frac12(K^I\diff L_I-L_I\diff K^I).  \label{omega-2}
\ee
The function $f$ in equation \eqref{5d-metric} takes the form,
\be
 f=(Z_1Z_2Z_3)^{-1/3}, 
\ee
where the three functions $Z_I$ are specified in terms of the harmonic functions as,
\be
Z_I = \frac{1}{2V} C_{IJK} K^J K^K+L_I.
\ee

The three vectors supporting the solution take the form, 
\be
 A^I =- \frac{1}{Z_I} (\diff t+k) + \frac{K^I}{V}(\diff z + A)+\xi^I + \diff t, 
\ee
with the three-dimensional one-forms $\xi^I$ given as,
\be	
\star_3\diff\xi^I=-\diff K^I.
\ee
Finally, the scalars supporting the solution are,
\be
 h^I = (fZ_I)^{-1}.
\ee

For the ACS saddle solution, the 8 harmonic functions  take the general form 
\be
H = h + \frac{\gamma_N}{r_N} + \frac{\gamma_S}{r_S}, 
\ee
where $r_N$ and $r_S$ are the distances from the points $\vec y_N$ and $\vec y_S$ in the three dimensional base metric, 
\bea
r_N &=& |\vec y - \vec y_N| = \sqrt{(y^1 - y^1_N)^2 + (y^2 - y^2_N)^2 + (y^3 - y^3_N)^2},   \label{north-pole-location-1} \\
r_S &=& |\vec y - \vec y_S| = \sqrt{(y^1 - y^1_S)^2 + (y^2 - y^2_S)^2 + (y^3 - y^3_S)^2}. \label{south-pole-location-1}
\eea
 The pole locations are, 
\begin{align}
\vec y_N &= \{y^1_N, y^2_N,y^3_N\} =\left\{0,0, -\frac i4 \sqrt{1-c^2} \, d\right\} \label{north-pole-location}, \\
\vec y_S &= \{y^1_S, y^2_S,y^3_S\} =\left\{ 0,0, \frac i4 \sqrt{1-c^2} \, d\right\} \label{south-pole-location}.
\end{align}
In terms of the $\{\rho,\theta,\varphi\}$ coordinates, $r_N,r_S$ take the form,
\begin{align}
	&r_N=\frac14(\rho^2+i\sqrt{1-c^2} \, d \cos 2\theta ),\\
	&r_S=\frac14(\rho^2-i\sqrt{1-c^2} \, d \cos 2\theta ).
\end{align}
Note that given our parameter restrictions  \eqref{restriction-1}--\eqref{restriction-2}, the pole locations are real. The 8 harmonic functions are,
\begin{align} 
&K^I = \frac{k^I_N}{r_N}+ \frac{k^I_S}{r_S}, &
&V = \frac{v_N}{r_N}+\frac{v_S}{r_S}, & \\
&L_I= 1+ \frac{l_{IN}}{r_N}+\frac{l_{IS}}{r_S}, & 
&M= \frac{m_N}{r_N}+\frac{m_S}{r_S}, &
\end{align}
where,
\begin{align}\label{coe-harmonic-1}
&v_N = (v_S)^* = \frac12\bigg(1+i\frac {J_{3L}}{D}\bigg) , \\
&k^I_N = (k^I_S)^*  = \frac i{2}\frac{Q^{(1)}Q^{(2)}Q^{(3)}}{\Qi D} , \\
&m_N = (m_S)^* = - \frac1{32} \left(J_{3L}+\frac{i}{2}\left(D - \frac{J_{3L}^2}{D}\right)\right), \\
&l_{IN} = (l_{IS})^*  = \frac{ 1}{8}\Qi \left(1-i \frac {J_{3L}}{D}\right),\label{coe-harmonic-4}
\end{align}
with $D$ given in \eqref{restriction-1}. These harmonic functions are our first key result of this paper. Having written the ACS solution in the Bena-Warner form, it follows that the solution is supersymmetric. The Killing spinors for the general Bena-Warner class of solutions were given in \cite{Bena:2004de}. We note that the Killing spinors constructed in \cite{Hegde:2023jmp} are also of the same form. The non-real nature of the solution does not introduce additional subtleties. 

\section{The attractor nature of the ACS saddle}

\label{sec:attractor}

In this section, we  show that the ACS saddle  exhibits the new form of attraction~\cite{Boruch:2023gfn}, i.e., the coefficients at the poles of the harmonic functions \eqref{coe-harmonic-1}--\eqref{coe-harmonic-4}  are determined by the solution to the attractor equations for the BMPV black hole charges.

To show this, it is useful to consider further (formal) dimensional reduction to
four dimensions. This gives the ${\cal N}=2$ STU model.  Four-dimensional STU model is governed by a prepotential function $F$ depending on four complex
scalars $X^M, M=0,1, 2,\,3$, 
\be
F(X) = - \frac{X^1 X^2 X^3}{X^0}.\label{preSTU}
\ee
For this model, the physical bosonic degrees of freedom are the metric, three complex scalars (six real scalars),
 and four one-forms $\check
A^M$. It is convenient to work with four complex scalars $X^M$ to begin with and later introduce a gauge fixing.

The 4D Lagrangian in a standard set of conventions can be obtained following the formalism reviewed in \cite{Mohaupt:2000mj}. In this formalism at several stages one needs to take the complex conjugation.  In the solution that we are considering  the metric and other fields also take complex values. 
In taking the complex conjugate, one should remember that this operation does not act on  the solution, i.e., even if the metric is complex, the complex conjugate of  $g_{\mu\nu}$ in the supergravity formalism give us back $g_{\mu\nu}$. We define,
$
F_M ={\partial F\over \partial X^M}, 
$
and work in the so-called $D$-gauge~\cite{Mohaupt:2000mj}, 
\be
 i (\bar{X}^M F_M - \bar{F}_M X^M ) = 1,
\ee
to ensure that the Einstein-Hilbert term is canonically normalised.

The relation between the  five-dimensional and the four-dimensional 
fields
is via dimensional reduction.
We write the five-dimensional metric  as, 
\begin{equation}
ds^2_5 = \wt f^{2}(\diff z - \check A^0)^2 + \wt f^{-1} \diff s^2_4, \label{sol-5d-1}
\end{equation}
and the vectors as,
\begin{equation}
 A^I = \chi^I(\diff z - \check A^0)+ \check A^I, \qquad I = 1, 2, 3. \label{sol-5d-2}
\end{equation}
Note the minus signs in front of $\check A^0$ terms. In writing these equations, we assume that  that no five-dimensional field depends on the coordinate $z$. As a result, there is a four-dimensional description. The dimensionally reduced Lagrangian reads, see, e.g., \cite{Virmani:2012kw},
\bea 
\mathcal{L}_4 &=&  R \star_4 \mathbf{1} - \frac{1}{2}G_{IJ} \star_4 \diff h^I \wedge
\diff h^J  - \frac{3}{2\wt f^2}\star_4 \diff \wt f \wedge \diff \wt f \nonumber  - \frac{\wt f^3}{2}\star_4
\check F^0 \wedge \check F^0 \\
&& - \frac{1}{2 \wt f^2}G_{IJ} \star_4 \diff  \chi^I \wedge \diff  \chi^J  - \frac{\wt f}{2}
G_{IJ} \star_4 (\check F^I -\chi^I \check F^0) \wedge (\check
F^J -\chi^J \check F^0) \label{STUaction}
\\
&&-\frac{1}{2}C_{IJK}\chi^I \check F^J  \wedge \check F^K  +
\frac{1}{2} C_{IJK}\chi^I \chi^J \check F^0  \wedge \check F^K - \frac{1}{6}
C_{IJK} \chi^I \chi^J \chi^K \check F^0 \wedge \check F^0\, .
 \nn
\eea
This Lagrangian is the STU model with the four-dimensional metric being $\diff s^2_4$ and the vectors being $ \check A^M = (\check A^0, \check A^I)$ and $\check F^M = \mathrm{d} \check A^M$. The scalars $\chi^I$ and $h^I$ combine to form the complex scalars in the STU model as 
 \be
z^I  = \frac{X^I}{X^0} =  \chi^I + i \wt f h^I, 
\ee 
in the D-gauge. 

For the class of solutions we are interested in, the four-dimensional metric takes the form
\be
\diff s^2 = - e^{2g} (\diff t + \omega)^2 + e^{-2 g} \diff s^2_{\mathrm{3d-base}},
\ee
where $\diff s^2_{\mathrm{3d-base}}$ is the metric on the three-dimensional flat base space and $\omega$ is a one-form on the three-dimensional base space\footnote{This $\omega$ and the $\omega$ introduced in \eqref{omega-1} and \eqref{omega-2} are the same. This will become clear shortly.}. Next, we introduce rescaled scalars $Y^M$. Here we are following the conventions and notation of \cite{Mohaupt:2000mj, 0009234}, which are slightly different from \cite{Boruch:2023gfn, 2402.03297}. We define,
\be
Y^M = e^{-g} \bar{h} X^M,
\ee
where $h$ is a position dependent phase that enter the description of the solutions. $h$ is also the phase of the central charge function $Z$, but we do not need this information in the following, $Z = e^{-g} h$. 
The homogeneity of the pre-potential function, 
\be
G(Y):= F(X (Y)),
\ee
implies, 
\be
G_M := \frac{\partial}{\partial Y^M} G(Y) = e^{-g} \bar{h} F_M.
\ee
The equations of motion of supergravity relate  $\{Y^M - \bar{Y}^M, G_M - \bar{G}_M\}$ to harmonic functions $\{H^M, H_M\}$ via the generalized stabilization equations \cite{9705169},
\bea
Y^M - \bar{Y}^M &=& i H^M, \label{GSE-1} \\
G_M - \bar{G}_M &=&  i H_M. \label{GSE-2}
\eea
These equations determine the full spatial dependence of $Y^M$ in terms of the harmonic functions. 

In terms of the harmonic functions, a rich class of half-BPS solutions of the STU model take the form~\cite{9705169}, 
\be
e^{-2g} = \Sigma = -i (Y^M \bar G_M - \bar{Y}^M G_M), 
\ee
such that
\bea
\Sigma^2 &=& - \left(\sum_{M=0}^{3}H^M H_M\right)^2 + \sum_{I, J, K, J', K' = 1}^{3} C_{IJK}H^J H^K C^{IJ'K'}H_{J'}H_{K'} + 4 H^0 H_1 H_2 H_3 
\nonumber  \\ 
& & 
- 4 H_0 H^1 H^2 H^3.
\eea
The physical scalars take the form,
\be
\chi^I = \frac{2 H^I H_I - \sum_{M=0}^{3}(H^M H_M)}{\sum_{J, K = 1}^{3}C_{IJK}H^J H^K + 2 H^0 H_I},
\ee
and
\be
\wt f h^I = \frac{\Sigma}{\sum_{J, K = 1}^{3}C_{IJK}H^J H^K + 2 H^0 H_I}.
\ee
The vectors take the form,
\be
\frac{1}{\sqrt{2}} \check A^M= - \partial_{H_M}(\log \Sigma) (\diff t + \omega) + A_3^M, \qquad \diff A_3^M = - \star_3 \diff  H^M. \label{AM-formula}
\ee
and finally the one-form $\omega$ is given as, 
\be
\star_3 \diff \omega = H^M \diff  H_M- H_M \diff H^M.
\ee
The relation between the 8 harmonic functions $\{H^M, H_M\}$ and the 8 harmonic functions that enter the description of the solutions in the Bena-Warner  form is as follows
\begin{align}\label{harmonic_normalisation1}
H^0 &= \frac{1}{\sqrt{2}} \, V \,, &H_0 &= \sqrt{2} \, M\, ,\\ 
H^1 &= \frac{1}{\sqrt{2}} \, K^1 \,, &H_1 &=  \frac{1}{\sqrt{2}} \, L^1\, ,\\
H^2 &= \frac{1}{\sqrt{2}} \, K^2\,, &H_2 &= \frac{1}{\sqrt{2}} \, L^2\, ,\\
H^3 &= \frac{1}{\sqrt{2}}  \, K^3\,, &H_3 &= \frac{1}{\sqrt{2}} \, L^3\,.  \label{harmonic_normalisation4}
\end{align}

For the spherically symmetric asymptotically flat  black hole in four-dimensions with charges $\{P^M, Q_M\}$ the harmonic functions take the form, 
\bea
H^M &=& \frac{P^M}{r} + \OO(1), \\
H_M &=& \frac{Q_M}{r} + \OO(1),
\eea
where $r = \sqrt{(y^1)^2+ (y^2)^2+ (y^3)^2}$ is the standard radial coordinate on the three-dimensional base and the horizon of the  spherically symmetric black hole is located at $r=0$.
The generalized stabilization equations \eqref{GSE-1}--\eqref{GSE-2} give rise to the 
attractor equations in the $r \to 0$ limit,
\bea
Y^M_* - \bar Y^M_* &=& i P^M,  \label{attractor-1} \\
G_M{}_* - \bar G_M{}_* &=& i Q_M,  \label{attractor-2}
\eea
where
\bea
Y^M &=& \frac{Y^M_*}{r} + \OO(1), \\
G_M &=& \frac{G_M{}_*}{r} + \OO(1), \\
\bar Y^M &=& \frac{\bar Y^M_*}{r} + \OO(1), \\
\bar G_M &=& \frac{\bar G_M{}_*}{r} + \OO(1),
\eea
as $r\to0$. Given the charges $\{P^M, Q_M\}$ the attractor equations \eqref{attractor-1}--\eqref{attractor-2} determine 
$\{ Y^M_*, \bar Y^M_*, G_M{}_*, \bar G_M{}_* \}.$ 
\subsection{The new form of attraction}

For the solutions we are interested in, the harmonic functions have only two poles, and they are taken as 
\be
H = h + \frac{\gamma_N}{r_N}  + \frac{\gamma_S}{r_S},
\ee
where $r_N$ and $r_S$ are the distances to the point $\{y^1, y^2, y^3\}$ from the 
north and the south poles, cf.~\eqref{north-pole-location-1}--\eqref{south-pole-location-1}.  The charges  $\gamma_N^M$, $\gamma_{NM}$, $\gamma_S^M$ and $\gamma_{SM}$ 
at the poles of the harmonic functions
are determined
from the equations \cite{Boruch:2023gfn}:
\bea\label{egammasum_1}
\gamma_N^M +  \gamma_S^M = P^M, \\
\gamma_{NM} +  \gamma_{SM} = Q_M\, , \label{egammasum_2}
\eea
where $P^M$ and $Q_M$ are the electric and magnetic charges carried by the
black hole, and by demanding that,
\bea 
Y^M  & =& {i \gamma_N^M\over r_N} + \OO(1), \\  
G_M & = & {i\gamma_{NM}\over r_N}  +\OO(1), \\
 \bar Y^M  & =&  \OO(1)\, , \\
\bar G_M &=& \OO(1)\, ,
\eea
as $r_N \to 0$, and 
\bea
   Y^M  &=& \OO(1), \\ 
 G_M &=& \OO(1)\, ,  \\
\bar Y^M  &=& 
-{i \gamma_S^M\over r_S} +\OO(1)\, , \\
\bar G_M &=&
-{i\gamma_{SM}\over r_S} +\OO(1)\, ,
\eea
as $r_S\to 0$.  In this way, the generalized stabilization equations \eqref{GSE-1}--\eqref{GSE-2} are satisfied at the poles.

It then follows from \eqref{egammasum_1}--\eqref{egammasum_2} and \eqref{attractor-1}--\eqref{attractor-2} that for the given charges $\{P^M, Q_N\}$, 
\be
\{i\gamma_N^M, i\gamma_{NM}, -i\gamma_S^M, -i\gamma_{SM}\}
=\{ Y^M_*,  G_M{}_*, \bar Y^M_*, \bar G_M{}_* \}. 
\ee
This implies, 
\be
\{\gamma_N^M, \gamma_{NM}, \gamma_S^M, \gamma_{SM}\} = \{-i Y^M_*, -i G_M{}_*, i\bar Y^M_*, i \bar G_M{}_*\}.
\ee
Moreover, since for the spherically symmetric black hole  $Y^M_*, G_M{}_*$ are related to $\bar Y^M_*, \bar G_M{}_*$  by complex conjugation, it also follows that 
\be
\gamma_N = (\gamma_S)^*.
\ee

\subsection{4D-5D connection and the harmonic functions}
The ACS saddle solution computes the supersymmetric index for the BMPV black hole with three independent charges. In order to obtain the coefficients at the poles for the harmonic functions, 
\be
\{\gamma_N^M, \gamma_{NM}, \gamma_S^M, \gamma_{SM}\},
\ee 
from the new attractor mechanism reviewed in the previous section, we need to view the 5D asymptotically flat BMPV black hole from the 4D perspective. The 4D-5D connection \cite{Gaiotto:2005gf} allows us to identify the 4D charges $\{P^M, Q_M\}$ for the 5D BMPV black hole. We have\footnote{The factors of $\sqrt{2}$ are there because there is a $\sqrt{2}$ in our eq.~\eqref{AM-formula}. The origin of this factor is the difference in the normalisation of the vector fields. In the $N=2$ supergravity formalism, the vectors are typically normalised  so that in the Lagrangian the kinetic term for a vector is $-\frac{1}{2} F_{\mu \nu}F^{\mu \nu}$  (see e.g., \cite{Mohaupt:2000mj, 0009234}) when the coefficient of the Einstein-Hilbert term is unity. In our case the vector kinetic terms are normalised (in a more standard manner) in eq.~\eqref{STUaction} as $-\frac{1}{4} F_{\mu \nu}F^{\mu \nu}$.}:
\begin{align}
P^0 &= \frac{1}{\sqrt{2}} \, , &  Q_0 &= - \frac{1}{8\sqrt{2}}J_{3L}\, , & \\
P^1 &= 0 \, , & Q_1 &= \frac{1}{4\sqrt{2}}Q^{(1)}\, ,& \\
P^2 &= 0\, , & Q_2 &= \frac{1}{4\sqrt{2}}Q^{(2)}\, ,&\\
P^3 &= 0\, , & Q_3 &= \frac{1}{4\sqrt{2}}Q^{(3)}\, .& 
\end{align}
The solution to the (spherically symmetric) attractor equations then gives, 
\begin{align}
\gamma_N^0 &= \frac{1}{2\sqrt{2}}\left(1 + i\frac{ J_{3L}}{D} \right)\,, &\gamma_{N0} &= -\frac{1}{16\sqrt{2}}\left(J_{3L} + \frac{i}{2} \left( D - \frac{J_{3L}^2}{D}\right)\right)\, ,\\
\gamma_N^1 &= \frac{i}{2\sqrt{2}} \frac{Q^{(2)}Q^{(3)}}{D}\,, &\gamma_{N1} &= \frac{1}{8 \sqrt{2}}Q^{(1)}\left(1- i\frac{J_{3L}}{D}\right)\, ,\\
\gamma_N^2 &= \frac{i}{2\sqrt{2}} \frac{Q^{(1)}Q^{(3)}}{D}\,, &\gamma_{N2} &=\frac{1}{8 \sqrt{2}}Q^{(2)}\left(1- i\frac{J_{3L}}{D}\right)\, ,\\
\gamma_N^3 &= \frac{i}{2\sqrt{2}} \frac{Q^{(1)}Q^{(2)}}{D}\,, &\gamma_{N3} &=\frac{1}{8 \sqrt{2}}Q^{(3)}\left(1- i\frac{J_{3L}}{D}\right)\, ,
\end{align}
where $D$ was defined in \eqref{restriction-1}. These expressions perfectly match with the coefficients of the harmonic functions computed in the previous section \eqref{coe-harmonic-1}--\eqref{coe-harmonic-4} upon taking into account the normalisation \eqref{harmonic_normalisation1}--\eqref{harmonic_normalisation4}. This matching is our second key result.

\section{Conclusions}
\label{sec:conclusions}

In this paper, we have written the Anupam-Chowdhury-Sen (ACS) non-extremal saddle solution \cite{Anupam:2023yns} in the Bena-Warner form that manifests its supersymmetric nature. Using the 4D-5D connection, we have also shown that the saddle solution exhibits the new form of attraction. Since for the corresponding extremal black holes the 4D-5D connection is well established~\cite{Gaiotto:2005gf}, it is perhaps not surprising the 4D-5D connection also allows us to relate the saddles that contribute to the 4D and 5D indices. 

  It is now well appreciated that the AdS$_2$/CFT$_1$ correspondence gives a  prescription for computing the degeneracy of black hole microstates in terms of the path integral over fields living in the near-horizon region. The near-horizon path integral receives contributions only from a class of configurations which are invariant under a subgroup of the supersymmetry transformations~\cite{Banerjee:2009af, Iliesiu:2022kny, Sen:2023dps}. Examples of such saddle points have been identified. Our current understanding of the saddle points of the path integral that uses asymptotically flat geometry to compute the supersymmetric index is not at the same level. In this paper, we have taken a small step towards improving our understanding of the saddle points, where we have constructed the Killing spinors and have established some key features related to the attractor mechanism for the dominant saddle that contributes to the 5D black hole index. We hope that in the future localisation techniques will allow to show that the path integral receives contributions from a special class of configurations;  a sub-class of them may be related to our construction in a direct way (e.g., as $\mathbb{Z}_k$ orbifolds). In this context, the results reported in this paper may be significant.

Our results offer other opportunities for future works, too. In refs.~\cite{2402.03297, Hegde:2024bmb}, the agreement between the entropy and index was shown to hold for half BPS black holes in $N=2$ supergravity with higher derivative terms in four-dimensions. The corresponding five-dimensional theories with higher derivative terms are well studied \cite{Hanaki:2006pj, Banerjee:2011ts}. As a future direction, it will be useful to extend the analysis of \cite{Hegde:2024bmb} to five-dimensions. We hope to report on this in our future work.

\bigskip

\noindent{\bf Acknowledgements:} 
We thank Subramanya Hegde, Ashoke Sen, and P Shanmugapriya for discussions. S.A. thanks CMI for hospitality during
the course of this work. P. D. acknowledges funding from the Institute of Eminence intitiative of the Government of India through IC\&SR of IIT Madras for the project ``SB22231259PHETWO008479''.  P. D. also thanks the Institute of Mathematical Sciences, Chennai where this work was done at the initial stages. The work of A.V. was partly supported by SERB Core Research Grant CRG/2023/000545 and by the ``Scholar in Residence'' program of IIT Gandhinagar.

\end{document}